\documentstyle[11pt,aaspp4]{article}

\begin{document}

\title{ASCA Discovery of an X-ray Pulsar in the Error Box of SGR1900+14}

\author{K. Hurley, P. Li}
\affil{University of California, Berkeley, Space Sciences Laboratory,
Berkeley, CA 94720-7450}
\authoremail{khurley@sunspot.ssl.berkeley.edu}
\author{C. Kouveliotou} 
\affil{Universities Space Research Association at NASA Marshall Space Flight Center, 
ES-84, Huntsville AL 35812}
\author{T. Murakami, M. Ando}
\affil{ISAS, 3-1-1 Yoshinodai, Sagamihara, Kanagawa, Japan 229}
\author{T. Strohmayer}
\affil{NASA Goddard Space Flight Center,
Greenbelt, MD 20771}
\author{J. van Paradijs\altaffilmark{1}, University of Alabama in Huntsville, AL 35899}
\author{F. Vrba, C. Luginbuhl}
\affil{U.S.Naval Observatory, Box 1149, Flagstaff Station, Flagstaff, AZ 86002-1149}
\author{A. Yoshida}
\affil{RIKEN, 2-1, Hirowawa, Wako, Saitama, 351-01, Japan}
\author{I. Smith}
\affil{Department of Space Physics and Astronomy, Rice University, MS-108, 6100 South
Main, Houston, TX 77005-1892}
\author{S. Kulkarni}
\affil{California Institute of Technology, Department of Astronomy, 105-24, 
Pasadena, CA 91125}

\altaffiltext{1}{Astronomical Institute `Anton Pannekoek',
University of Amsterdam, The Netherlands}

\begin{abstract}

We present a 2 - 10 keV ASCA observation of the field around the soft gamma
repeater SGR1900+14.  One quiescent X-ray source was detected in this observation, and it
was in the SGR error box.  In 2 - 10 keV X-rays, its spectrum may be fit by a power law with
index -2.2, and its unabsorbed flux is $\rm9.6 x 10^{-12} erg \ cm^{-2}\ s^{-1}$.  We
also find a clear 5.16 s period.  The properties of the three well-studied soft gamma
repeaters are remarkably similar to one another, and provide evidence that all of
them are associated with young, strongly magnetized neutron stars in supernova
remnants.

\end{abstract}

\keywords{gamma rays: bursts --- stars: neutron --- X-rays: stars --- 
supernova remnants}

\section{Introduction}

Four soft gamma repeaters (SGRs) are now known to exist with certainty, and there is good 
evidence that all of them are associated
with supernova remnants (SNRs).  The position of SGR0525-66 is 
consistent with that of the N49 supernova remnant in the Large Magellanic Cloud  
(Cline et al. 1982).  SGR1806-20 (Atteia et al. 1987) is associated with the  
SNR G10.0-0.3, located towards the galactic center (Kulkarni \& 
Frail 1993; Kouveliotou et al. 1994; Kulkarni et al. 
1994; Murakami et al. 1994).  SGR1627-41 may be associated with
G337.0-0.1 (Woods et al., 1998; Hurley et al. 1998a,b).  SGR1900+14 lies
close to G42.8+0.6 (Kouveliotou et al. 1994; Hurley et al. 1998c).  Three of the four are quiescent
soft X-ray point sources: SGR0525-66 (Rothschild, Kulkarni, \& Lingenfelter 1994),
SGR1806-20 (Murakami et al. 1994), and SGR1900+14 (Hurley et al. 1994, 1998c).  Finally
two of the SGRs, 0525-66 (Mazets et al. 1979, Barat et al. 1979) and 1900+14 (Hurley et
al. 1998d) display  periodicities in their bursting emission, and one, 1806-20
(Kouveliotou et al. 1998a) in its quiescent
emission, indicating that they are neutron stars.

In this paper we present soft X-ray observations of the quiescent X-ray source 
associated with SGR1900+14 by the Advanced
Satellite for Cosmology and Astrophysics (ASCA) which indicate that it, too
displays a periodicity in its quiescent emission.  Furthermore, its X-ray spectrum is similar to 
that of SGR1806-20.  In another paper
(Hurley et al. 1998c) we present evidence, based on burst localization 
by the interplanetary network, that this
X-ray source is indeed associated with the SGR.  Thus SGR1900+14 is similar to
the other SGRs in many respects, pointing towards a unified model
for these sources.

Kouveliotou et al. (1994) suggested that SGR1900+14 might be associated with
one of two SNRs, G43.9+1.6 or G42.8+0.6.  Both the ROSAT sky survey (Vasisht et al. 1994) and a pointed observation 
at the position of G42.8+0.6 (Hurley et al. 1996) indicated that a point-like quiescent X-ray source was present near this SNR.  This source was also
associated  with one of the two network synthesis error boxes for this SGR (Hurley et al. 1994).  No ROSAT source was detected 
in the second network synthesis error box (Li et al. 1997), and the more recent
localization for SGR1900+14 (Hurley et al. 1998c) indicates that the first
error box is indeed the one which contains the SGR and the quiescent X-ray source.  Optical and infrared observations of the X-ray position 
revealed a peculiar double M star system within the X-ray source error box (Vrba et al. 1996).    A rough estimate of the probability of finding a random X-ray source of any intensity
in the error box for this SGR may be obtained from the statistics of
the WGA catalog (White, Giommi, and Angelini 1995).  There are $\rm \sim 10,000$ sources
located at galactic latitudes $\rm \vert b \vert < 20^{o}$, or $\rm 1.12x10^{-3}$ sources/
sq. arcmin., and the error box area is $\rm \sim$ 1.6 sq. arcmin. (Hurley et al. 1998c).  Thus the
probability is $\rm \sim$ 0.0018.  This estimate makes no correction for the sky exposure
as a function of galactic latitude, and encompasses the widely varying sensitivities
of the ROSAT pointed observations in the public domain (typical observation times
5000 - 30,000 s).  However, as we show below, due to the fact that the period of
the quiescent source is identical to that found in the giant flare of 1998 August 27
(Hurley et al. 1998d), the association between the SGR and the X-ray source does
not depend strongly on this probability.

\section{ASCA Observations}

The position of the quiescent X-ray point source associated with SGR1900+14,
$\alpha(2000)=19^{\rm h} 07^{\rm m} 14^{\rm s}, \delta(2000)=9^{\rm o} 19 \arcmin 19 \arcsec$, 
was known from previous ROSAT observations (Hurley et al. 1994).  
ASCA was pointed to within $\sim 0.1^{\rm o}$ of this position on 1998 April 30.
(An earlier attempt at an observation on 1997 October 22 was incorrectly pointed). 
The observation began around 20:55 UT and lasted 2.3 days, resulting in
74 ksec of on-source exposure with the Solid State Imaging Spectrometer (SIS), and 84.6 ksec
with the Gas Imaging Spectrometer (GIS).  We used the standard screening criteria for such parameters as Earth elevation angle, South Atlantic
Anomaly, and cutoff rigidity to extract photons.  These are described in the
ASCA Data Reduction Guide, Version 2 \footnote{available on-line at http://heasarc.gsfc.nasa.gov/docs/asca/abc/abc.html}.
No bursts from the source
were observed by \it Ulysses \rm, BATSE, KONUS, or ASCA during the observation.
One quiescent source was detected, at $\alpha(2000)=19^{\rm h} 07^{\rm m} 16.4^{\rm s},
\delta(2000)=+09^{\rm o} 19 \arcmin 44.1 \arcsec$ with an error radius 40 $\arcsec$, consistent with the
position of the ROSAT source.  The source location was
determined with the Ximage source detection tool, which uses a sliding window
technique.

The region used for spectral analysis consisted of a 6 $\arcmin$
radius circle centered at the source position; background was taken from the
same observation, using a
6 $\arcmin$ radius circle at a region where no source was present, as determined
by the same sliding window technique.
Spectral fitting was done using XSPEC and three trial functions: blackbody, thermal
bremsstrahlung, and a power law, all with absorption.  Using GIS2 and GIS3 data,
a power law gave the best fit, with a reduced $\chi^2$ of 1.48 for 182 degrees of freedom.  
The best reduced $\chi^2$ for the blackbody fit was 4.38, and that for the
bremsstrahlung was 2.04.  We therefore adopt the power law fit;
the best fit photon power law index
was 2.25 $\pm$ 0.04, and the absorption was $\rm n_H=(2.16 \pm 0.07)x10^{22} cm^{-2}$.
This corresponds to an unabsorbed 2 - 10 keV flux of $(\rm9.6 \pm0.7) x 10^{-12} erg \ cm^{-2} \ s^{-1}$. 
The fit is shown in figure 1.  

Previously, the distance to SGR1900+14 had been estimated as 5 kpc using the $\Sigma$-D
relation (e.g. Vasisht et al. 1994).  The measurement of n$_H$ now permits an
independent estimate.  Using the W3nH tool available on-line from the HEASARC
\footnote{http://heasarc.gsfc.nasa.gov/docs/frames/mb\_w3browse.html} we obtain
$\approx$5.7 kpc.  Both methods are subject to uncertainties; in what follows, we adopt
a distance of 5 kpc. 

To search for periodicity, light curves were constructed with 0.5 s binning from
the sum of the GIS2 and GIS3 data, by extracting $\approx$ 1 - 10 keV counts from a 4 $\arcmin$ radius
circular region around the
source.  The power spectrum is shown in figure 2.  A significant
barycenter-corrected period was found at 5.1589715 $\pm$ 0.0000008 s.  The distribution of powers closely follows
a $\chi^2$ distribution with two degrees of freedom .  Taking into
account the number of frequencies searched, the chance probability of finding a peak this
strong or stronger in the power spectrum may be estimated at $\rm 4.7 x 10^{-14}$,
a conservative number which does not take the presence of the harmonics into account.  A folded
light curve is shown in figure 3.  This period agrees well with that found for the SGR
in outburst on 1998 August 27 (Hurley et al. 1998d), making the identification of the
quiescent source and the SGR quite secure.

\section{Discussion and Conclusion}

The ASCA observations of SGR1900+14 may be used to estimate the
magnetic field of the neutron star in two ways.  First, we can
use the association between the quiescent X-ray source and G42.8+0.6. 
The observable lifetime
of SNRs is less than $\sim$ 20,000 years (Braun, Goss, and Lyne 1989).  The
characteristic age $\tau$ of a rotating, magnetized neutron star is related
to the period P and its derivative $\dot{P}$ by $\tau=P/2\dot{P}$,
while P and $\dot{P}$ are related to the field strength B by
P$\dot{P}$=AB$^2$, where A=9.8x10$^{-40}$ s G$^{-2}$ (e.g. Lyne, Manchester, and Taylor
1985).  Thus
$B=\sqrt{P^2/2A\tau}$.  For SGR1900+14, we obtain B=2x10$^{14}$G for an
age of 10$^4$ y, remarkably
similar to estimates made from the energy contained in the soft tail of
the giant flare of 1998 August 27 (Hurley et al. 1998d) and from direct measurement
of $\dot{P}$ (Kouveliotou et al. 1998b).  It is also comparable to the field strength
of SGR1806-20, estimated from direct measurements of P and $\dot{P}$ for
that source (Kouveliotou et al. 1998a).  This suggests that SGR behavior can be described by the magnetar
model (Thompson and Duncan 1995).  If this is the case, a second
estimate of B comes from one of the predictions of that model, namely that
persistent soft X-ray emission is powered by crustal magnetic energy:
$\frac{B_{crust}^2}{8\pi}4\pi R_*^2 \Delta R \gtrsim L_x \tau$,
where R$_*$ is the radius of the neutron star (10 km), $\Delta R$ is the thickness
of the crust ($\sim$ 1 km), and we take L$_x$=2.7x10$^{34}$ erg s$^{-1}$ for
a distance of 5 kpc and $\tau$=10,000 y.  We obtain
$B_{crust} \gtrsim 4 x 10^{14} G$.  The two estimates would agree (B $\sim$
2.5x10$^{14}$ G) for an age
$\tau \sim 7100 $y, or about 70\% of the lifetime of a typical magnetar, in the
model.

Table 1 compares the properties of the three SGRs and their soft X-ray counterparts.
The X-ray spectrum of SGR0525 is unknown.  However, the spectra of SGR1806 and
SGR1900 have identical power law indices; their 2 - 10 keV luminosities agree
to within an order of magnitude when their estimated distances are taken into
account.  The periods of all three sources agree to within a factor of $\rm \sim
1.6$.  The four known SGRs therefore share many similarities -- in their periods, the
properties of their
quiescent soft X-ray sources, their ages, their magnetic fields, and their associations with supernova remnants.  This provides further
evidence that, as a class, the SGRs are strongly magnetized young neutron stars whose
behavior can be described by the magnetar model of Thompson and Duncan (1995).

\acknowledgments
KH and PL are grateful to NASA for support under the ASCA AO-6  
Guest Investigator Program.  CK acknowledges
support under the CGRO guest investigator program (NAG 5-2560).
We are also grateful to Dr. K. Mitsuda who kindly provided his
computer code to confirm the period.

\newpage

\begin{figure}
\figurenum{1}
\plotone{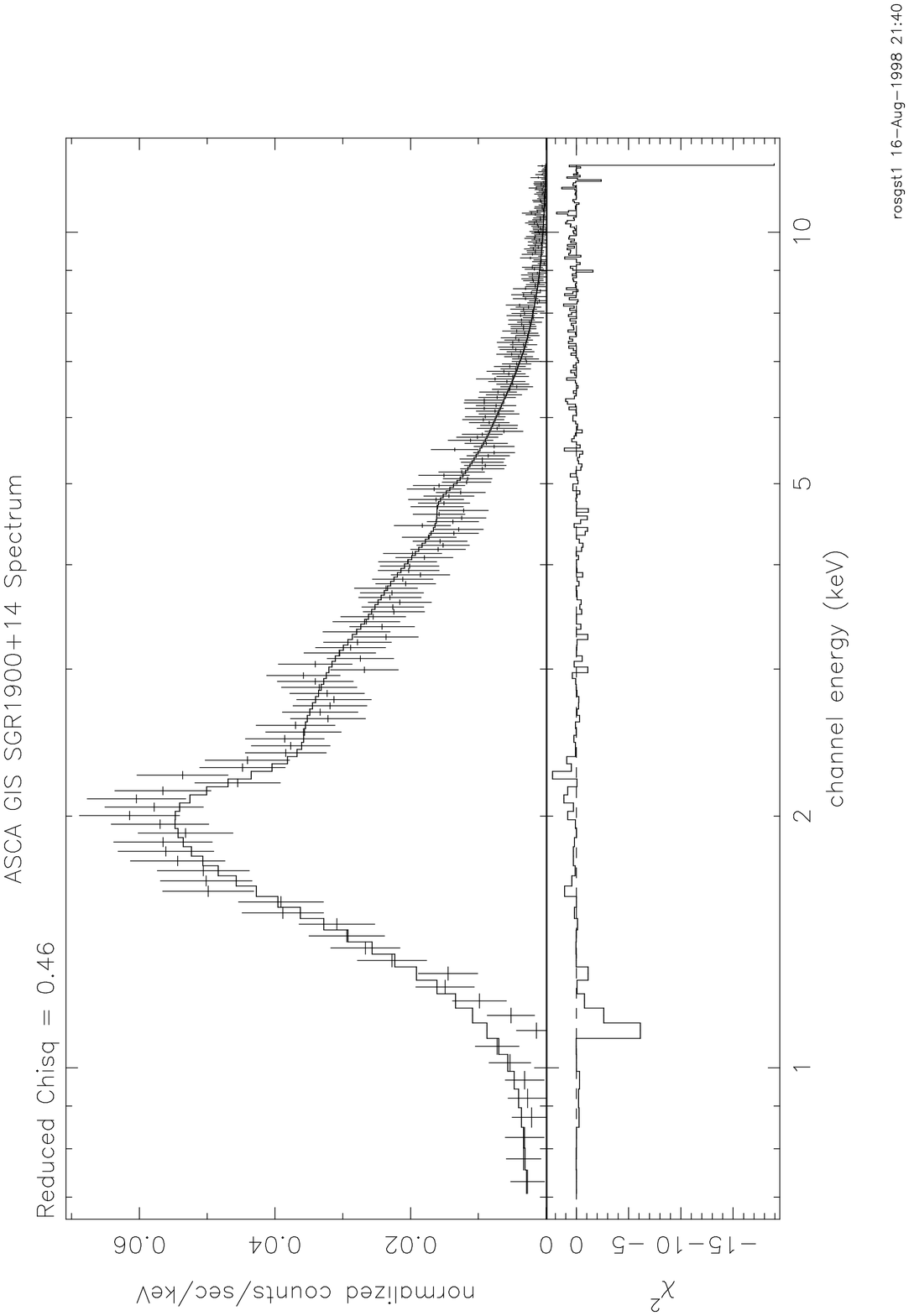}
\caption{Best fit power law spectrum and residuals for the GIS2 and GIS3 data.
The large residuals around 1 keV are believed to be caused by a small gain
shift. }
\end{figure}

\begin{figure}
\figurenum{2}
\plotone{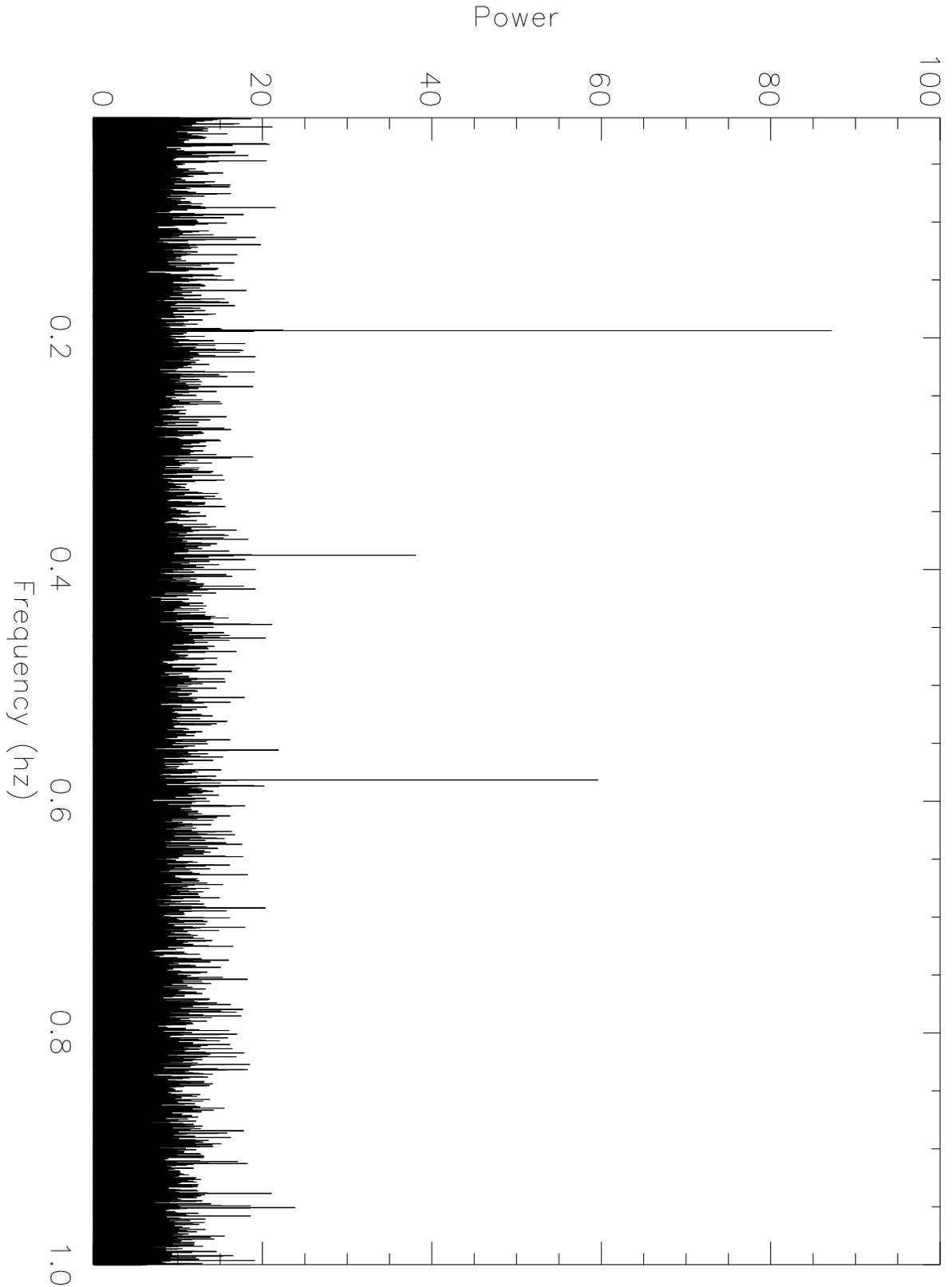}
\caption{Power spectrum of the 2 - 10 keV soft X-ray source associated with
SGR1900+14.   The 0.19384-Hz pulsed signal and its first and second harmonics
are well above the noise level.}
\end{figure}

\begin{figure}
\figurenum{3}
\plotone{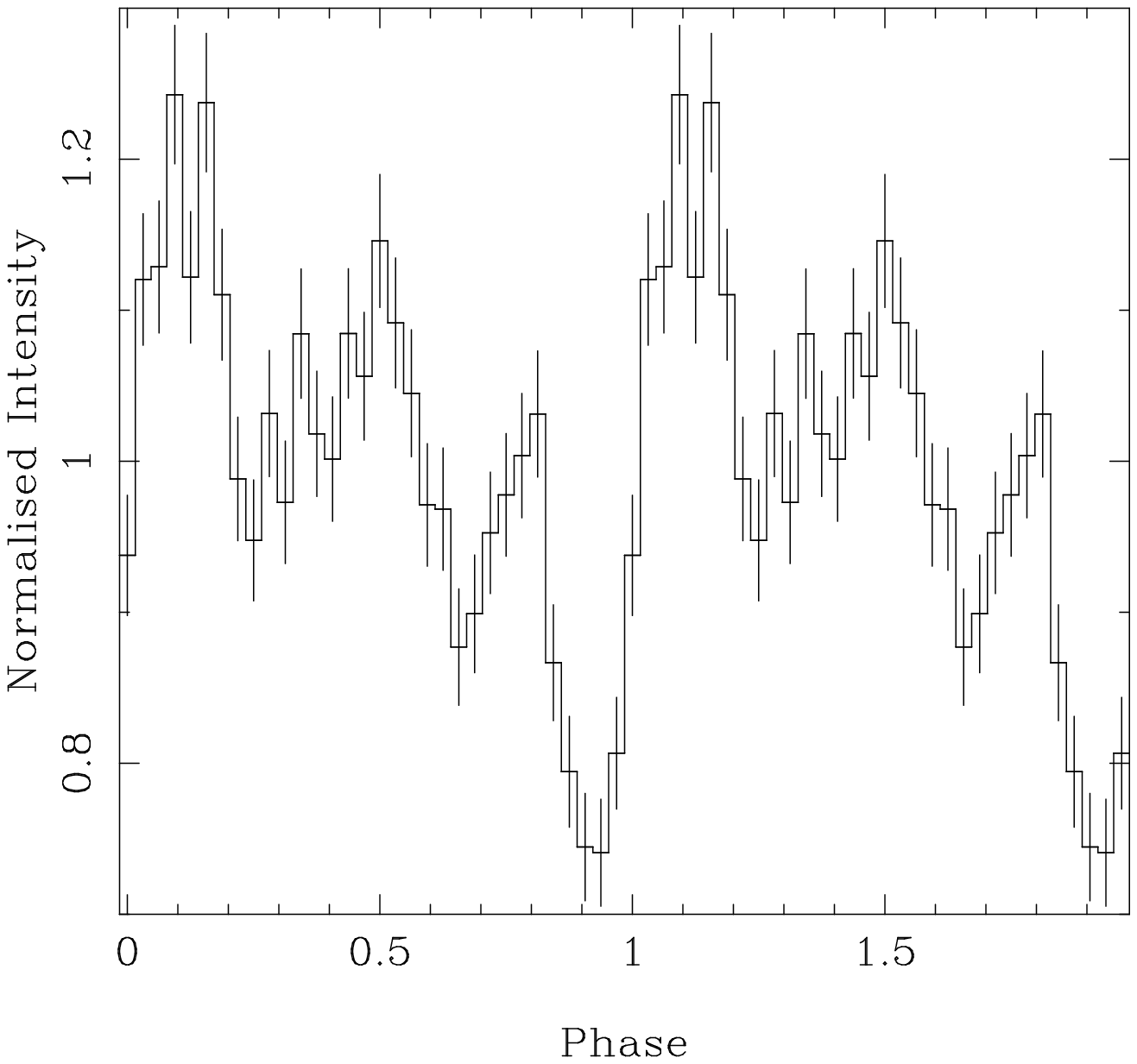}
\caption{Epoch folded pulse profile of SGR1900+14 from the 2 - 10 keV ASCA data.}
\end{figure}

\clearpage
\begin{deluxetable}{ccccc}
\tablecaption{\it Properties of three SGRs}
%\tablewidth{0pt}
\tablehead{
\colhead{SGR}& \colhead{Soft X-ray spectrum}  & \colhead{Flux, $\rm erg \ cm^{-2} \ s^{-1}$} & \colhead{Period, s} & \colhead{Distance, kpc}
}
\startdata
0525-66\tablenotemark{1,2,3,4} & ?\tablenotemark{5} & $\rm 2x10^{-12}$ & 8   & 55 \nl
1806-20\tablenotemark{2,3,6}   & $\rm E^{-2.2}$ & $\rm 1x10^{-11}$   & 7.5 & $\rm \sim 15(?)$ \nl
1900+14                        & $\rm E^{-2.2}$ & $\rm 9.6x10^{-12}$ & 5.2 & $\rm \sim 5(?)$  \nl                 

\enddata
\tablenotetext{1}{Rothschild, Kulkarni, and Lingenfelter 1994}
\tablenotetext{2}{Murakami 1995}
\tablenotetext{3}{Murakami et al. 1996}
\tablenotetext{4}{Barat et al. 1979}
\tablenotetext{5}{Source was detected by the ROSAT HRI, but not resolved by ASCA; no spectral
measurement available.  Flux is for 0.1 - 2.4 keV}
\tablenotetext{6}{Kouveliotou et al. 1998a}
\end{deluxetable}

\end{document}